\documentclass[%
 reprint,
 amsmath,amssymb,
 aps,
prb,
]{revtex4-1}

\usepackage{graphicx}
\usepackage{dcolumn}
\usepackage{bm}
\usepackage{upgreek} 
\usepackage{dcolumn}
\usepackage{bm}
\usepackage{textcomp}
\usepackage{siunitx}
\mathchardef\mhyphen="2D
\usepackage{soul,xcolor}
\usepackage{color}

\setul{1ex}{0.4ex}

\begin{document}
\setstcolor{red}

\title{Detection of spin pumping from YIG by spin-charge conversion in a Au\textbar Ni$_{80}$Fe$_{20}$ spin-valve structure}

\author{N. Vlietstra}
\affiliation{Physics of Nanodevices, Zernike Institute for Advanced Materials, University of Groningen, Groningen, The Netherlands}

\author{F. K. Dejene}
\affiliation{Max Planck Institute for Microstructure Physics, Weinberg 2, 06120 Halle(Saale), Germany}

\author{B. J. van Wees}
\affiliation{Physics of Nanodevices, Zernike Institute for Advanced Materials, University of Groningen, Groningen, The Netherlands}

\date{\today}

\begin{abstract}

Many experiments have shown the detection of spin-currents driven by radio-frequency spin pumping from yttrium iron garnet (YIG), by making use of the inverse spin-Hall effect, which is present in materials with strong spin-orbit coupling, such as Pt. Here we show that it is also possible to directly detect the resonance-driven spin-current using Au\textbar permalloy (Py, Ni$_{80}$Fe$_{20}$) devices, where Py is used as a detector for the spins pumped across the YIG\textbar Au interface. This detection mechanism is equivalent to the spin-current detection in metallic non-local spin-valve devices. By finite element modeling we compare the pumped spin-current from a reference Pt strip with the detected signals from the Au\textbar Py devices. We find that for one series of Au\textbar Py devices the calculated spin pumping signals mostly match the measurements, within 20\%, whereas for a second series of devices additional signals are present which are up to a factor 10 higher than the calculated signals from spin pumping. We also identify contributions from thermoelectric effects caused by the resonant (spin-related) and non-resonant heating of the YIG. Thermocouples are used to investigate the presence of these thermal effects and to quantify the magnitude of the Spin-(dependent-)Seebeck effect. Several additional features are observed, which are also discussed. 

\begin{description}

\item[PACS numbers]
72.25.-b, 75.78.-n, 76.50.+g, 85.75.-d


\end{description}
\end{abstract}


\maketitle

\section{Introduction}
Employing a ferro/ferrimagnetic insulating material (FMI) for spintronics research has attracted a lot of interest in the past years owing to the possibility of generating pure spin-currents, without accompanying spurious charge-currents. Besides, in these materials, it is shown that spin information can be transported over large distances on the $\upmu$m-scale \citep{Ludo} or even mm-scale \citep{Kajiwara2010nature,Magnonics}, opening up new possibilities for spin-based data storage and transport. In these devices, Yttrium iron garnet (YIG), which is a room-temperature FMI with very low magnetic damping, is most often employed. Together with the (inverse) spin-Hall effect ((I)SHE) in Pt, it offers a platform for studying pure spin-current generation, transport and detection. An example of such an experiment is the electrical detection of spin pumping in a YIG\textbar Pt system, where the resonance of the YIG magnetization leads to a spin-current pumped into the adjacent Pt layer, which can electrically be detected via the ISHE. \citep{Kajiwara2010nature,VariousSizeAndo2011,SaitohFreqDep,CastelPRB,SpinPump}

Pure spin-currents can also be generated and detected by making use of metallic magnetic\textbar non-magnetic nanostructures such as permalloy (Ni$_{80}$Fe$_{20}$, Py) or cobalt.\citep{Jedema} This method is mostly used in spin-valve structures, where a spin-current is generated by sending a charge-current through one magnet, which can be detected (either locally or non-locally) by a second magnetic strip, as a change in electric potential when switching between relative parallel and anti-parallel magnetic states of both magnets.\citep{Fert,Grunberg} 

In this manuscript, we show that by combining a FMI (YIG) with a conducting magnetic material (Py) it is possible to electrically detect the magnetic resonance of the FMI, without the need of a high spin-orbit coupling material like Pt. Here the magnetic-resonance-induced dc spin-current pumped into an adjacent Au layer is detected as an electrical voltage by a Py detector connected to a Au spacer. 

This alternative method for detection of spin-currents from FMI-materials opens up new ways of investigating the origin of the spin-Seebeck effect\citep{TheoryMagnon} without the possible presence of non-equilibrium proximity magnetization in the heavy metal Pt.\citep{ChienPRL2012,XMCDchien,XMCDgoennenwein} Besides, because of its analogy to measuring a conventional spin-valve structure, this method also helps to determine the sign of the pumped spin-current from YIG into Pt,\citep{sign} and expands the possibilities for designing devices, including spin transport through FMI materials. 

In the experiments we first induce magnetic resonance in the YIG by sending RF currents through a microwave stripline, which is placed near the Au\textbar Py devices that are connected in series to maximize the total signal. 

Part of the build-up potential we attribute to the spin-current generation by spin pumping from the YIG into the adjacent structure [schematically shown in Fig. \ref{fig:Fig81}(a)]. Hereby we compare spin pumping signals from a standard YIG\textbar Pt device structure with the signal from YIG\textbar Au\textbar Py devices placed in series. Furthermore, we also identify signals that are related to heating and induction effects, which are rather small to explain the observed signals.

It is found that we not only detect the resonance spin pumping from the magnetic YIG layer, but also observe the Py resonance state. This self-detection of FMR by a Py strip has been observed before,\citep{Costache2} however, here we discuss that the mechanism is possibly different and related to the interaction of spins at the YIG interface.


\section{Sample characteristics}

The studied devices are fabricated on a $4 \times 4$ mm$^2$ sized sample, which is cut from a wafer consisting of a 500-$\upmu$m-thick single crystal (111)Gd$_3$Ga$_5$O$_{12}$ (GGG) substrate and a 210 nm thick layer of YIG, grown by liquid phase epitaxy (from the company Matesy GmbH). The YIG magnetization shows isotropic behavior of the magnetization in the film plane, with a low coercive field of less than 1 mT (measured by SQUID).

\begin{figure}[]
	\includegraphics[width=8.5cm]{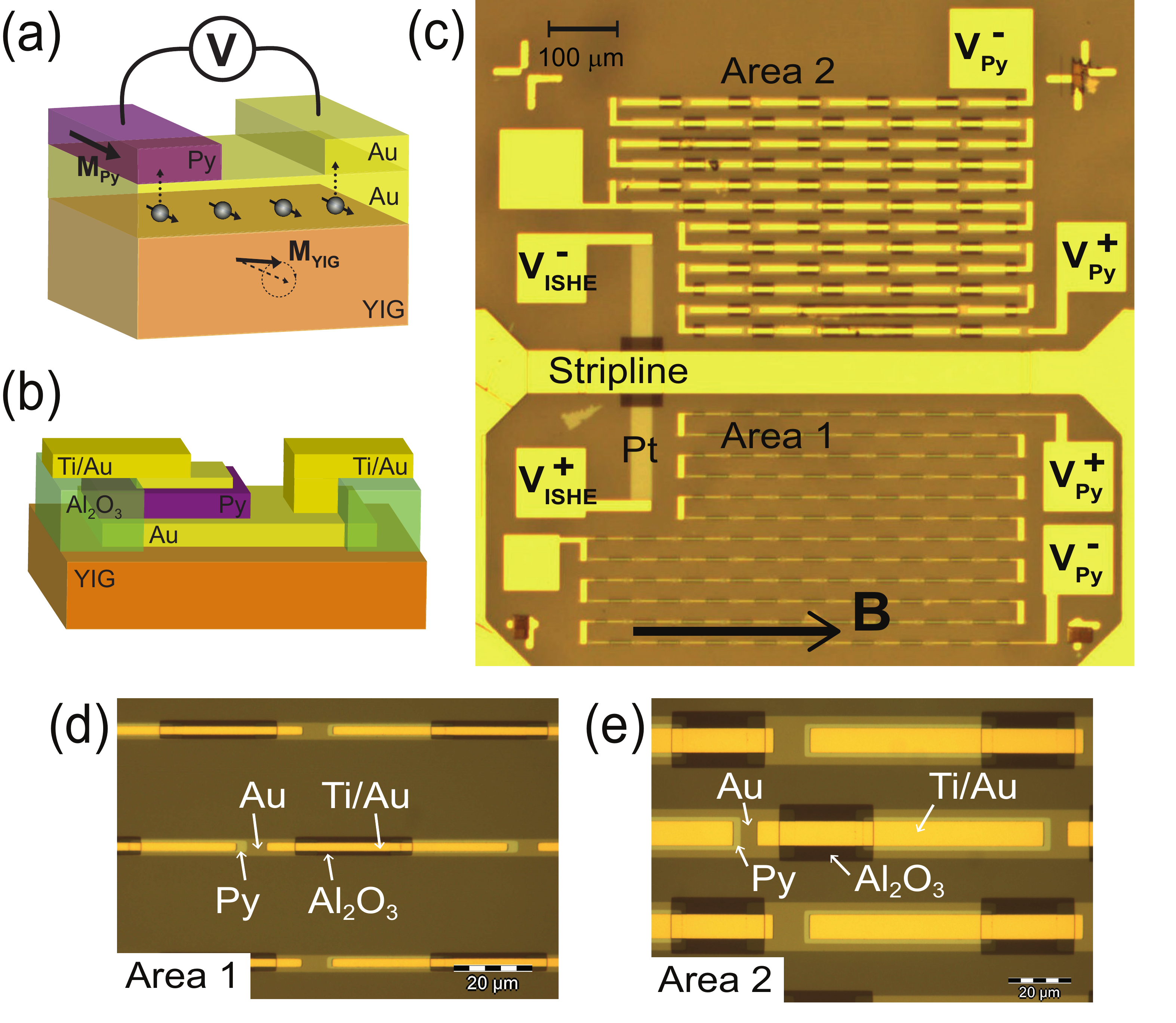}
	\caption{\label{fig:Fig81} 
		(a) Schematic representation of the spin pumping process and voltage detection in a YIG\textbar Au\textbar Py device. (b) Schematic drawing of one YIG\textbar Au\textbar Py device. Each device consists of Au (8 nm), Py (20 nm), Al$_2$O$_3$ (60 nm), and Ti\textbar Au (5\textbar 100 nm) layers. (c) Microscope image of the final device structures. In area 1 (area 2) 96 (62) YIG\textbar Au\textbar Py devices are placed in series. The Pt strip is used as a reference for the measurements on the YIG\textbar Au\textbar Py devices, and this strip is electrically insulated from the stripline by an Al$_2$O$_3$ layer (60 nm thick). During the measurements, an external magnetic field is applied and three separate sets of voltage probes are connected as pointed out in the figure. (d) and (e) show close-ups of the devices placed in area 1 and 2, respectively. The different material layers are marked.
	}
\end{figure}

Fig. \ref{fig:Fig81}(b) shows a schematic of one device from the studied series, fabricated by several steps of electron beam lithography. It consists of an 8-nm-thick Au layer deposited on YIG by dc sputtering, followed by a 20-nm-thick Py layer ($30 \times 2.5$ $\upmu$m$^2$ for area 1, and $60 \times 10$ $\upmu$m$^2$ for area 2), contacted with a top Ti\textbar Au layer of 5\textbar 100 nm, both deposited by e-beam evaporation. To prevent shorting between the top and bottom Au layers, when placing several devices in series, a 60-nm-thick Al$_2$O$_3$ layer was deposited over the edges of the Au\textbar Py stack before the deposition of the top Ti\textbar Au layer. Ar-ion milling has been used to clean the surfaces and etch the native oxide layer of Py before deposition of the Py and Ti\textbar Au layers, respectively. Fig. \ref{fig:Fig81}(c) shows a microscope image of a full series of devices. With the employed meander structure for the series of devices possible signals generated by the ISHE cancel out. In Fig. \ref{fig:Fig81}(c) also the reference Pt-strip ($400 \times 30$ $\upmu$m$^2$, 7-nm-thick, dc sputtered) can be seen, placed below and perpendicular to the 60 $\upmu$m wide Ti\textbar Au microwave stripline (5\textbar 100 nm thick) used to excite the magnetization resonance in the magnetic layers. Between the Pt-strip and the microwave stripline a 60-nm-thick Al$_2$O$_3$ layer (dark brown) is added, to prevent electrical shorts. Fig. \ref{fig:Fig81}(d) and \ref{fig:Fig81}(e) show a close-up of the devices in area 1 and 2, respectively.

For the spin pumping experiment using the YIG\textbar Pt system, the obtained signal scales linearly with the length of the Pt detection strip. For the YIG\textbar Au\textbar Py devices the detected signal is not directly scalable by the size of the device, rather by the number of YIG\textbar Au\textbar Py devices connected in series. As the expected signal for one YIG\textbar Au\textbar Py device is below the signal-to-noise ratio of our measurement setup, we fabricated a structure where we increase the detected signal by placing many separate devices in series. Two sets of devices were investigated, having different surface area, as is shown in fig. \ref{fig:Fig81}(c). For the presented experiments 96 (area 1) and 62 (area 2) YIG\textbar Au\textbar Py devices in series were used. 

\section{Measurement methods}
The microwave signal is generated by a Rohde-Schwarz vector network analyzer (ZVA-40), connected to the waveguide on the sample via a picoprobe GS microwave probe. To be able to use the low-noise lock-in detection method, the power of the applied microwave signal is modulated between `on' and `off' using a triggering signal which is synchronized with the trigger of the lock-in amplifiers. Typically used `on' and `off' powers are 10 dBm and -30 dBm, respectively. The RF frequency is fixed for each measurement (ranging from 1 GHz to 10 GHz), while sweeping the static in-plane magnetic field.

During this magnetic field sweep, the voltage from the Pt strip and both series of Py devices are separately recorded by connecting them to three different lock-in amplifiers, using the connections as is shown in Fig. \ref{fig:Fig81}(c). The ISHE voltage detected from the Pt strip is used as a reference for the data obtained from the Py devices. As the power is modulated during all measurements, the absorbed power cannot be measured simultaneously, and therefore no detailed information is obtained about the dependence of power absorption on applied RF frequency. The frequency dependence of the absorbed power was obtained by separate measurements of the S$_{11}$ parameter as in Ref. \citep{CastelPRB}. 
All measurements are performed at room temperature.

\section{Results and discussion}

\begin{figure}[tb]
	\includegraphics[width=8.5cm]{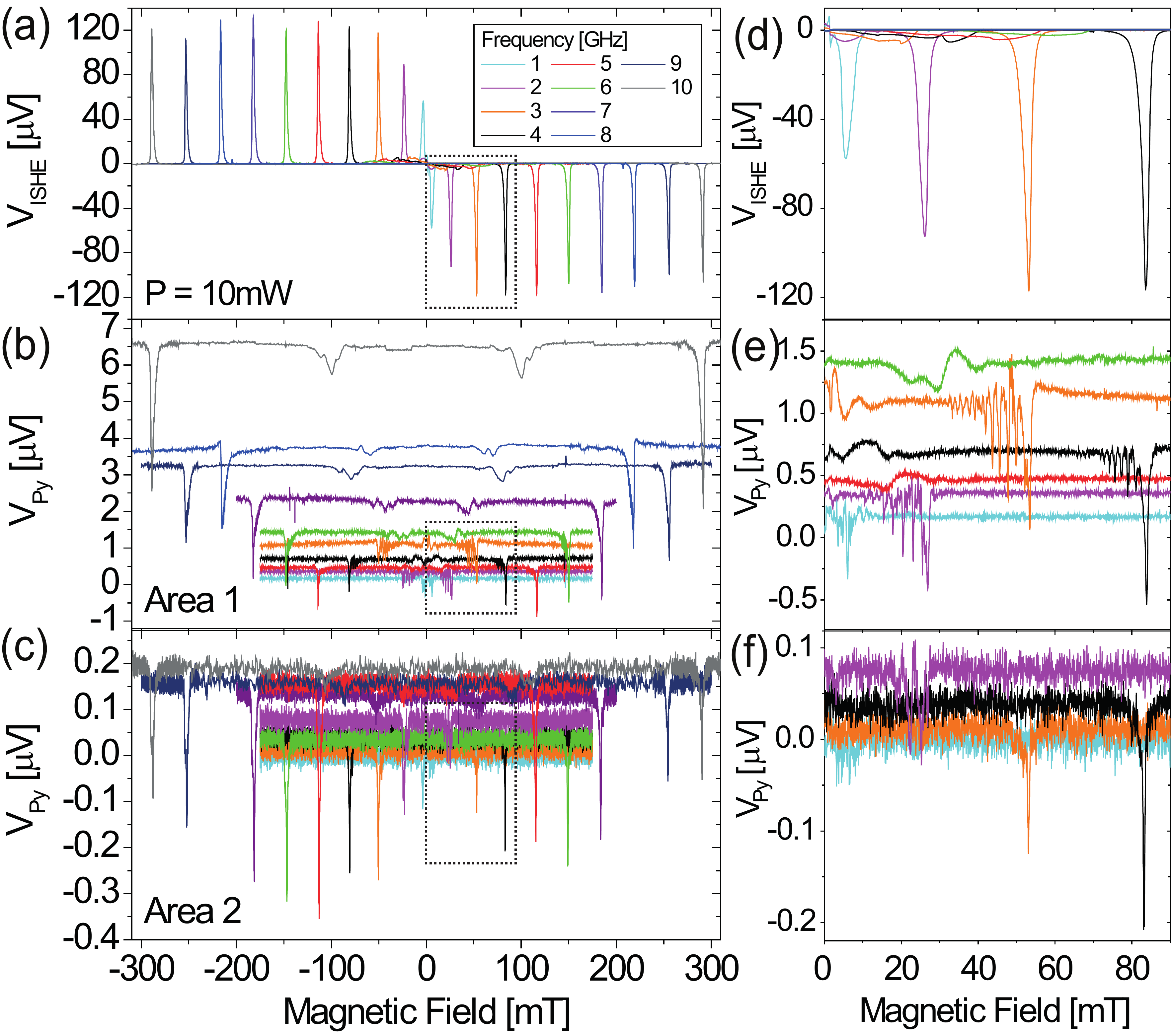}
	\caption{\label{fig:Fig82} 
		Magnetic field sweeps for different applied frequencies, with an applied RF power of 10 mW for (a) the Pt strip, detecting the ISHE voltage and (b), (c) voltage generated by the series of Py-devices in area 1 and 2, respectively. (d)-(f) close-up of marked areas in (a)-(c).
	}
\end{figure}

Results of the magnetic field sweeps for different RF frequencies are shown in Fig. \ref{fig:Fig82} where the ISHE voltage signal detected by the Pt strip [Fig.~\ref{fig:Fig82}(a)] and the corresponding signals from the series of Py devices in area 1 [Fig.~\ref{fig:Fig82}(b)] and area 2 [Fig.~\ref{fig:Fig82}(c)] are shown. The detected voltage of the Pt strip shows the expected peaks for YIG resonance, changing sign when changing the magnetic field direction, as also observed in for example Refs. \citep{VariousSizeAndo2011,CastelPRB,VincentYIG}. The magnitude of the peaks is in the order of 100 $\upmu$V. Comparing these results to the data shown in Figs. \ref{fig:Fig82}(b) and \ref{fig:Fig82}(c); peaks at exactly the same position are observed for the Py devices. Here the detected peaks do not change sign by changing the magnetic field direction, as the sign of the signal in these devices is determined by the relative orientation of the YIG and Py magnetization. Because of the low coercive fields ($<$10 mT) of both the YIG and Py layers, their magnetizations always align parallel to each other when applying a small magnetic field.

\begin{figure}[b]
	\includegraphics[width=8.5cm]{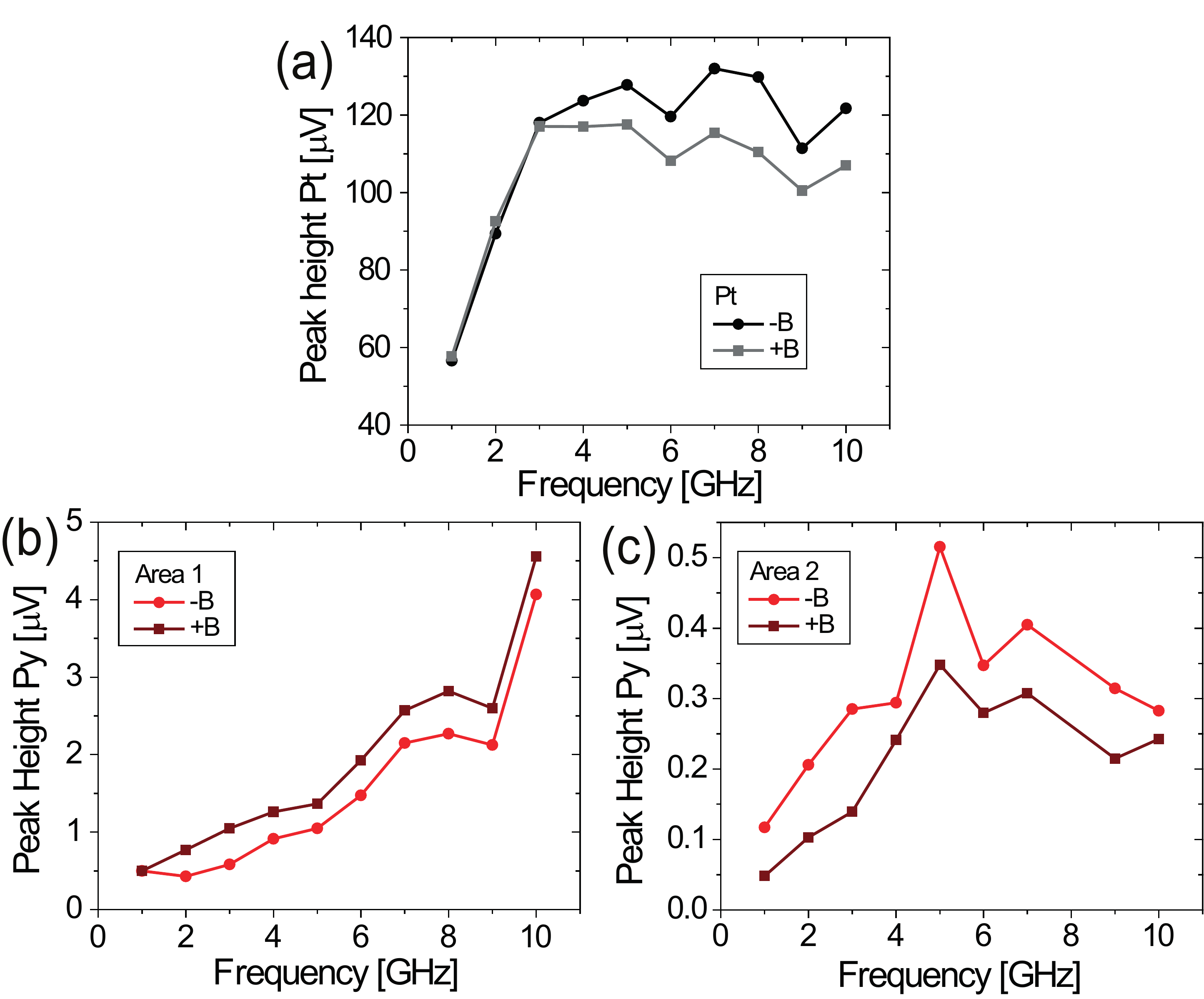}
	\caption{\label{fig:Fig823} 
		Magnitude of the peaks at YIG resonance obtained from the measurements shown in Fig. \ref{fig:Fig82}, for positive and negative applied magnetic fields as a function of applied RF frequency. (a) For the Pt strip, (b) and (c) for the Py devices of area 1 and 2, respectively.
	}
\end{figure}

At first glance, it appears that when the YIG is excited into resonance conditions, the pumped spin-current into the Au layer is detected as an electrical voltage generated by a conducting ferromagnet placed on top of the Au layer. Fig. \ref{fig:Fig823} shows the frequency dependence of the magnitude of the observed peaks at YIG resonance for all three types of devices (Pt strip and Py devices area 1 and 2). A few points can be made regarding these dependencies. First, where the dependence of the Pt shows some analogy with the dependence observed in area 2, the frequency dependence of the signals from area 1 (narrow Py strips) and area 2 (wider Py strips) largely differs. Besides, the signals from area 1 and 2 show about one order difference in magnitude. These observations show that by changing the surface area of the YIG\textbar Au\textbar Py devices, different physics phenomena can be present. In section \ref{section:calculation} we calculate the contribution of spin pumping to the observed signals in both Py device areas and discuss these results. 

Secondly, from Fig. \ref{fig:Fig823}, we observe that the signals for positive and negative applied magnetic fields consistently differ. Interestingly, for area 1 the higher signals are observed for positive applied fields, whereas for area 2 the higher signals appear for negative applied fields. At first glance, the device is fully symmetric and therefore one would not expect any dependence on the sign of the applied magnetic field. Further investigation leads to the existence of nonreciprocal magnetostatic surface spin-wave modes (MSSW), whose traveling direction (perpendicular to the stripline) is determined by the applied magnetic field direction.\citep{MSSWdirection,agrawal_microwave-induced_2014} As the spin pumping process in insensitive to the spin-wave mode, and the device areas 1 and 2 are placed on opposite sides of the stripline, the presence of these unidirectional MSSWs can lead to the observed difference in peak-height for positive and negative applied fields. 



Besides the peaks at YIG resonance, the Py devices show more peaks at lower applied magnetic fields (most clearly visible in area 1, Fig. \ref{fig:Fig82}(b)), these peaks are attributed to the ferromagnetic resonance of the magnetic Py layers, as will be shown in section \ref{section:peaks} and further discussed in section \ref{section:discussion}.

\subsection{Position of the resonance peaks}
\label{section:peaks}
\begin{figure}[b]
	\includegraphics[width=7cm]{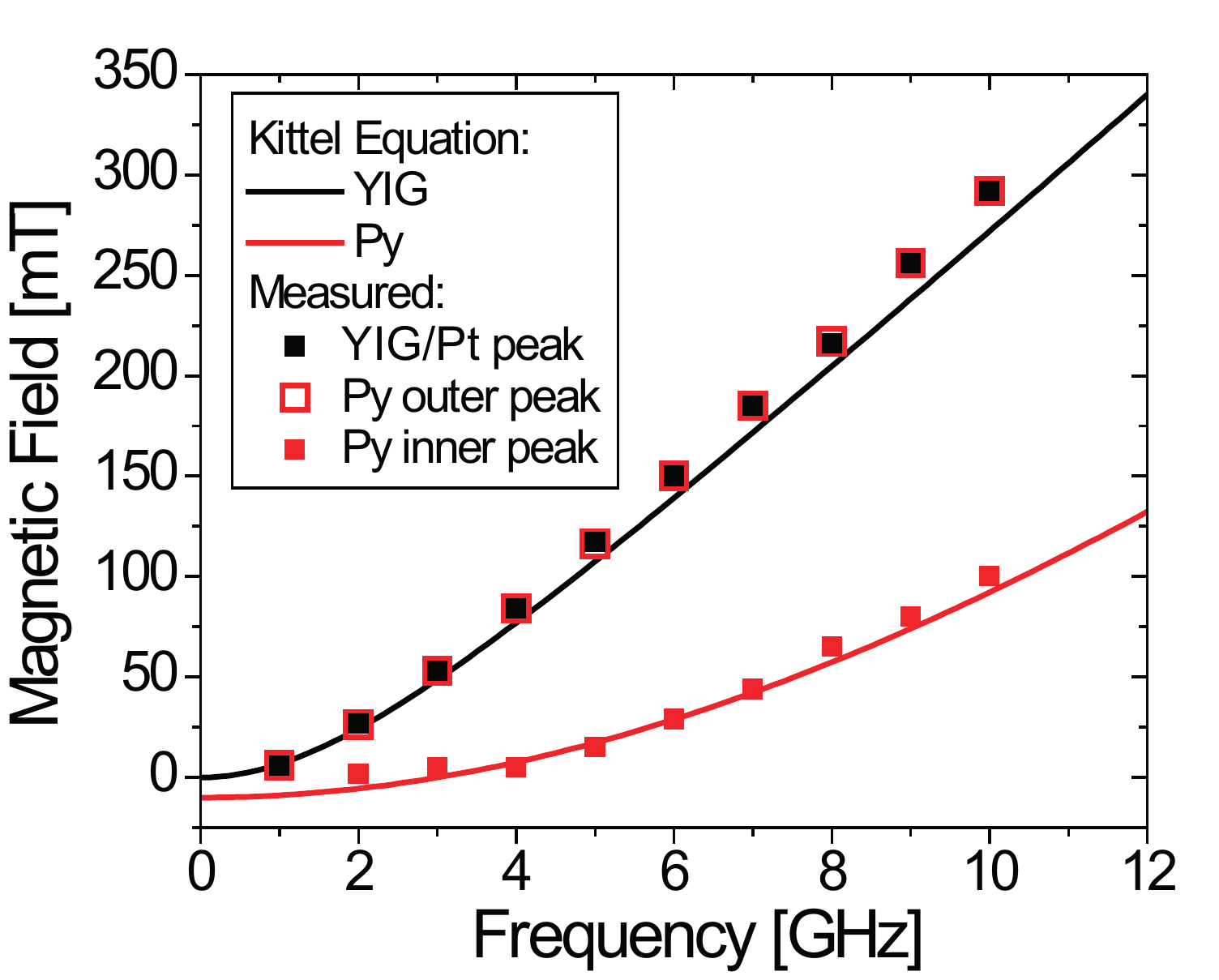}
	\caption{\label{fig:Fig83} 
		The measured peak positions for both the Py devices area 1 (red symbols) as well as the Pt strip (black symbols), plotted together with the calculated resonance conditions by the Kittel equation [Eqs. (\ref{eq:Kittel}) and (\ref{eq:KittelSimple})].
	}
\end{figure}

To check the position of the observed peaks with respect to the predicted resonance conditions of the magnetic layers (YIG and Py), the Kittel equation is used:\citep{Kittel,Kittel2}

\begin{equation}
\label{eq:Kittel}
f= \frac{\gamma}{2\pi}\sqrt{(B+N_\parallel \mu_0 M_s)(B+N_\perp \mu_0 M_s)},
\end{equation}
where $\gamma$ is the gyromagnetic ratio ($\gamma=176$ GHz/T), $B$ is the applied magnetic field, $N_\parallel$ and $N_\perp$ are in-plane and out-of-plane demagnetization factors and $\mu_0 M_s$ is the saturation magnetization. Taking $N_\parallel \mu_0 M_s=10$ mT and $N_\perp \mu_0 M_s=1.1$ T (consistent with previously reported values for Py)\citep{Costache,Costache2}, we obtain the red solid curve shown in Fig. \ref{fig:Fig83}, and find that the position of the inner peaks, detected only for the series of Py devices, matches the Py resonance conditions (the shown data is obtained from area 1). Therefore, the origin of the inner peaks is related to the Py layers being in ferromagnetic resonance.

The position of the outer peaks, detected for both series of Py devices, corresponds to the voltage peaks appearing in the Pt strip, which are ascribed to the YIG magnetization being in resonance. To check the expected YIG resonance conditions, a more simplified form of the Kittel equation can be used, because of its isotropic in-plane magnetization behavior ($N_\parallel=0$ and $N_\perp=1$):

\begin{equation}
\label{eq:KittelSimple}
f= \frac{\gamma}{2\pi}\sqrt{(B(B+\mu_0 M_s)}.
\end{equation}

For the curve corresponding to the YIG resonance peaks as shown in Fig. \ref{fig:Fig83}, $\mu_0 M_s=176$ mT is used, which is the reported bulk saturation magnetization of YIG.\citep{Kajiwara2010nature,Kurebayashi2011nmat,CastelPRB} The close match of the calculated curve and the measured data proves that the outer peaks from the Py devices and the peaks from the Pt strip indeed originate from the YIG magnetization being in resonance. 

\subsection{Estimation of the spin pumping signal in YIG\textbar Au\textbar Py devices}	
\label{section:calculation}
In this section we will only focus on the origin of the peaks at YIG resonance. The possible origin of the detected inner peaks for the Py devices will be discussed in section \ref{section:discussion}. Using the data from the Pt strip as a reference, we calculate the expected signal caused by spin pumping in the YIG\textbar Au\textbar Py devices. Different steps in this calculation are: 1) calculate the pumped spin-current density from the ISHE voltage detected by the Pt strip, 2) obtain an estimate for the spin-mixing conductance of the YIG\textbar Au interface, and 3) use a finite element spin transport model to find the expected spin pumping signal in the YIG\textbar Au\textbar Py devices, setting the results of 1) and 2) as boundary conditions and input parameters, respectively. For the calculations we assume that the spin accumulation in the layer adjacent to the YIG, defined as the ratio between the injected spin-current $J_s$ and the real part of the spin-mixing conductance $G_r$, is constant when considering different types of interfaces and devices.\citep{ImprovedGr}

\begin{figure}[]
	\includegraphics[width=6cm]{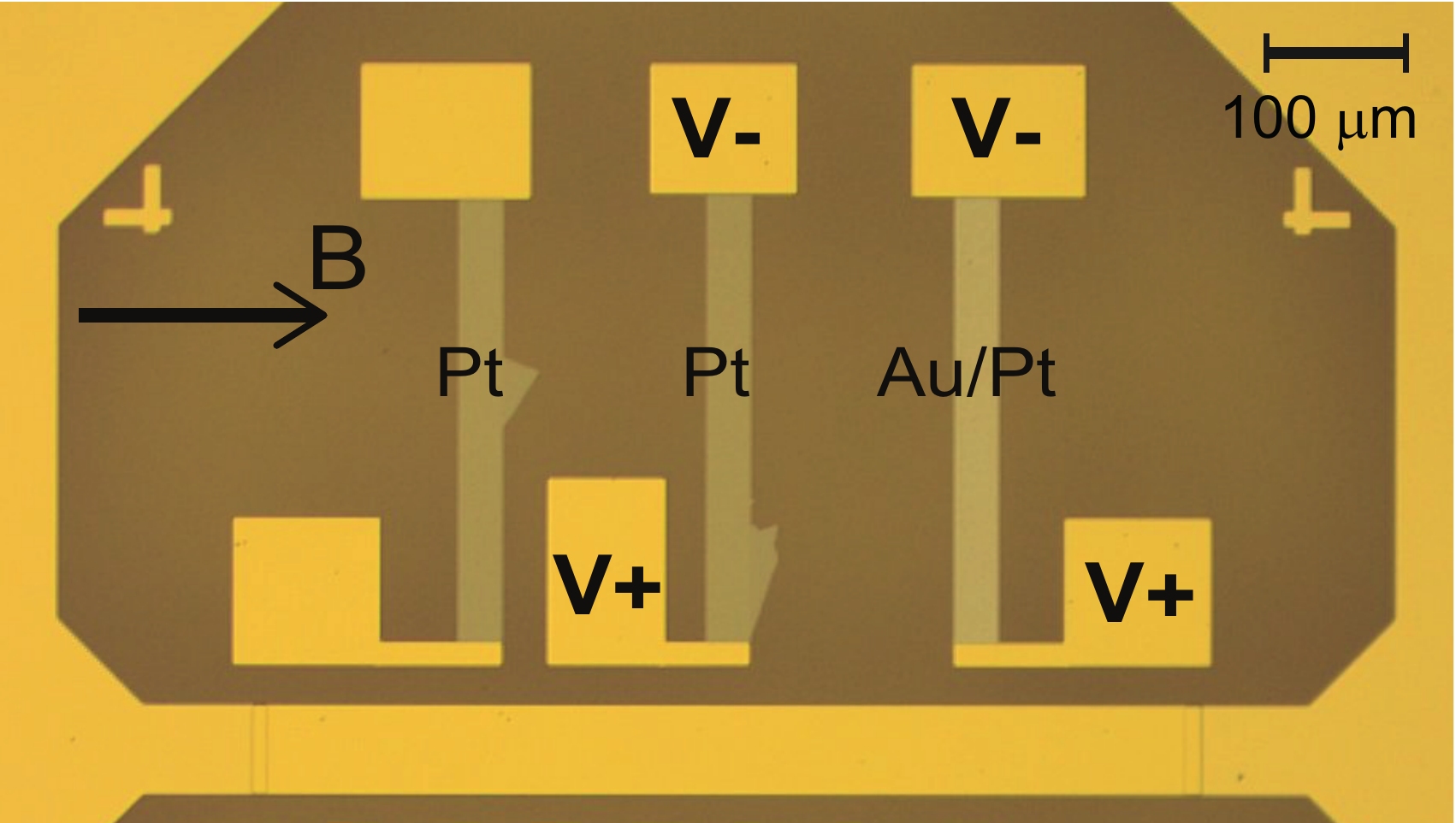}
	\caption{\label{fig:Fig834} 
		Microscope image of the Pt and Au\textbar Pt strips used to estimate the injected spin-currentby spin pumping at the location of the Py-devices.
	}
\end{figure}

As the magnitude of the excited spin-waves decays with distance from the stripline, we fabricated another sample, where a 6-nm-thick Pt strip and a Au\textbar Pt strip (8\textbar 6 nm) were placed exactly on the location of the series of Py devices in the previous batch, as is shown in Fig. \ref{fig:Fig834} (The dimensions of these strips are $334 \times 30$ $\upmu$m). Using this sample, the average injected spin-current on the location of the Py devices can directly be calculated. Signals obtained from these new devices show similar behavior as the data shown in Fig. \ref{fig:Fig82}(a) and Fig. \ref{fig:Fig823}(a), only the magnitude of the signals differs: The Pt (Au\textbar Pt) strip on the new sample results in ISHE-voltages around 30 $\upmu$V (3 $\upmu$V), compared to 120 $\upmu$V for the Pt strip directly below the RF line. These significantly lower signals prove the decay of spin-waves with distance from the stripline. The signal for the Au\textbar Pt strip is even more suppressed compared to the Pt strip, as the Au layer short-circuits the structure because of its lower resistance compared to Pt, while the inverse spin-Hall voltage is mainly generated in the Pt layer. In the following calculation, the Au\textbar Pt strip is modeled using the 3D finite-element modeling software Comsol Multiphysics, taking into account these losses. 

Step 1: From the measured ISHE-voltage $V_\textrm{ISHE}$ in the Pt strip, the injected spin-current density can be calculated by	
\begin{equation}
\label{eq:Js}
J_s = V_\textrm{ISHE} \frac{t_\textrm{Pt}}{l \rho \lambda \theta_\textrm{SH}} \frac{1}{\eta} \frac{1}{\tanh(\frac{t_\textrm{Pt}}{2\lambda})},
\end{equation}
where $t_\textrm{Pt}$, $l$, $\rho$, $\lambda$, and $\theta_\textrm{SH}$ are the thickness (6 nm), length (330 $\upmu$m), resistivity ($3.5\times10^{-7}$ $\Omega$m), spin relaxation length (1.2 nm) and the spin-Hall angle (0.08) of the Pt strip, respectively, and $\eta=[1+G_{r \textrm{,Pt}}\rho\lambda\coth(\frac{t_\textrm{Pt}}{2\lambda})]^{-1}$ is the backflow term,\citep{SSETheory} where $G_{r \textrm{,Pt}}$ is the spin-mixing conductance of the YIG\textbar Pt interface ($4.4\times10^{14}$ $\Omega^{-1}$m$^{-2}$). The given material properties are taken from previously reported spin-Hall magnetoresistance measurements.\citep{VlietstraSMR2} Using Eq. (\ref{eq:Js}), an injected spin-current density of $2.0\times10^7$ A/m$^2$ is found for a detected $V_\textrm{ISHE}$ of 30 $\upmu$V. 

Step 2: The spin-mixing conductance of the YIG\textbar Pt interface differs from the YIG\textbar Au interface, as present in the measured YIG\textbar Au\textbar Py devices. Therefore we use the obtained signals from the combined Au\textbar Pt strip to find an estimate of the spin-mixing conductance of the YIG\textbar Au interface $G_{r \textrm{,Au}}$. Spin pumping into the Au\textbar Pt strip results in ISHE signals in the order of 3 $\upmu$V. Using Comsol Multiphysics, we model the YIG\textbar Au\textbar Pt device, including spin-diffusion by the two-channel model and the spin-Hall effect as explained in Ref. \citep{Comsol}. To include the contribution of the spin-mixing conductance in this model, such that backflow is accounted for, a thin interface layer ($t=1$ nm) is defined between the YIG and Au layers (resulting in a stack: YIG\textbar interface\textbar Au\textbar Pt). The interface layer acts as an extra resistive channel for the injected spin-current, parallel to the spin-resistance of the device on top (in this case the Au\textbar Pt strip), such that there effectively are two spin-channels: one for backflow and one for injection into the Au layer. The conductivity of this interface layer is defined as $\sigma_\textrm{int}=G_{r \textrm{,Au}} \cdot t$. The input parameter at the interface\textbar Au boundary is the dc spin-current $J_s$ obtained in step 1 for the YIG\textbar Pt device. By scaling $J_s$ with the ratio of $G_{r\textrm{,Au}}$ and $G_{r\textrm{,Pt}}$ we take into account that the injected spin-current is lower when the spin-mixing conductance is lower. We now tune the value of $G_{r \textrm{,Au}}$ in the model such that the modeling result matches the measured $V_\textrm{ISHE}$ of the Au\textbar Pt strip. By doing so we find $G_{r \textrm{,Au}}=(2.2 \pm 0.2) \times10^{14}$ $\Omega^{-1}$m$^{-2}$, in order to match the measured $V_\textrm{ISHE}$ of the Au\textbar Pt strip. This value is similar as reported by Heinrich \textit{et al.},\citep{HeinrichInterface} who obtained $G_{r \textrm{,Au}}$-values up to $1.9 \times 10^{14}$ $\Omega^{-1}$m$^{-2}$. Used modeling parameters for the Au layer are $\sigma_\textrm{Au}=6.8\times10^6$ S/m and $\lambda_\textrm{Au}=80$ nm.\citep{Fasil} The modeling parameters for the Pt layer are as mentioned above.  

By replacing the Pt layer by the Au\textbar Pt strip, we find that the backflow 
spin-current almost doubles (around 75\% increase). This increased backflow is mainly caused by the larger spin-diffusion length in Au as compared to Pt, resulting in a higher spin-resistance for the injection of spins into the thin Au layer, as compared to the backflow spin-channel (In other words: Pt is a better spin-sink). Furthermore, the initially injected spin-current is a factor 2 lower for the YIG\textbar Au\textbar Pt strip, compared to YIG\textbar Pt, caused by the lower spin-mixing conductance. These two cases result in intrinsically lower signals when placing Au on top of YIG as compared to Pt. 

Step 3: After having calculated the injected spin-current and the spin-mixing conductance of the YIG\textbar Au interface, these parameters are used as input for the Comsol Multiphysics model of one YIG\textbar Au\textbar Py device. This model is again based on the two-channel model for spin transport, including an additional interface layer to add backflow to the model, as described above. Detailed information about the modeling and the used equations can be found in Ref. \citep{Comsol}. The different sizes of devices present in area 1 and area 2 are both separately modeled. 

The properties of the Py layer added to the model are $P=0.3$ (defining the spin polarization), $\sigma_\textrm{Py}=2.9\times10^6$ S/m and $\lambda_\textrm{Py}=5$ nm.\citep{Fasil} The spin-current density obtained from Eq. (\ref{eq:Js}) is used as input, and is set as a boundary flux/source term at the bottom interface of the Au layer. As explained in step 2, also here a thin interface layer is placed below the Au layer, such that it acts as a spin-current channel parallel to the injection of spin-current into the device. The above estimated value for $G_{r \textrm{,Au}}$ is used to define the interface layer conductivity. Also from this model we find a large backflow 
from the initially injected $J_s$, which is mainly caused by the spins accumulating at the Au\textbar Py interface, increasing the spin-resistance in the injection-channel with respect to the backflow spin-channel.

\begin{figure}[]
	\includegraphics[width=7cm]{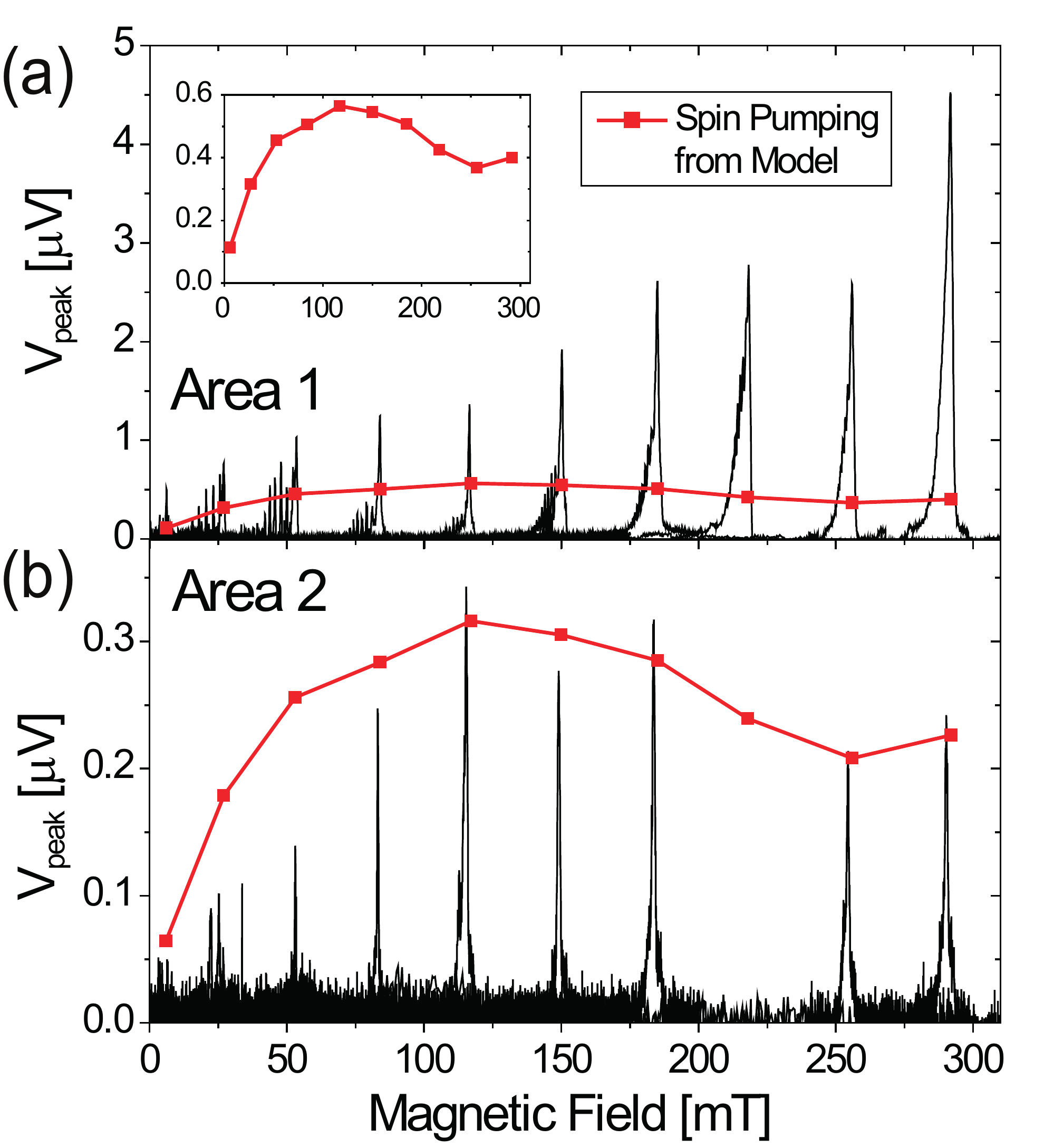}
	\caption{\label{fig:Fig84} 
		Calculated spin pumping voltage (red squares) versus measured signal (black peaks) as a function of applied RF frequency and magnetic field. The peaks from left to right correspond to applied RF frequencies from 1 to 10 GHz, respectively, directly copied from the measurements shown in Fig \ref{fig:Fig82}. (a) Results for the Py-devices in area 1. The inset shows a close-up of the calculated spin pumping voltage. (b) Results for the Py-devices in area 2.
	}
\end{figure}

To obtain the dependence on RF frequency of the expected voltage signal, $J_s$ was calculated from Eq. (\ref{eq:Js}) for each frequency, using the measurements on the YIG\textbar Pt strip, and the YIG\textbar Au\textbar Py model was run for each obtained $J_s$. The calculated voltage signal caused by spin pumping for one Py device was multiplied by 96 (62), the number of devices placed in series in area 1 (area 2), and the final results of these calculations are shown as the red curve in Fig. \ref{fig:Fig84}(a) and \ref{fig:Fig84}(b), together with the absolute values of the measured peaks, shown in black, for area 1 and 2, respectively.

For the Py-devices in area 1, the calculated spin pumping voltages are one order of magnitude smaller than the measured signals, from which we conclude that the major contribution of the measured signal is caused by another source than spin accumulation created by spin pumping. Additionally, the calculated signals generated by spin pumping clearly show a different dependence on frequency, as compared to the measured signals from the YIG\textbar Au\textbar Py devices. This also indicates that besides spin pumping there are other phenomena present which induce voltage signals in our devices. 

Interestingly the peaks obtained from the devices in area 2 show a very good agreement with the calculated spin pumping signals. From these results we conclude that in this case the major contribution of the measured signals is caused by spin pumping. 

Comparing the results from area 1 and area 2, it is clear that in the experiments the exact device geometry largely influences the signal and even results in a totally different dependence on applied RF frequency. From the calculations of the spin pumping signals, such a big change in behavior cannot be reproduced and therefore additional phenomena must be present and becoming more prominent for more narrow devices, as present in area 1. 

As thermoelectric effects might also play a role in the performed experiments, next section discusses some further investigation of possible signals related to RF-heating.

\subsection{Thermal effects}
While an RF current flows through the stripline, heat is absorbed by the YIG layer causing the YIG temperature to rise. Together with Eddy currents that are induced in the Py devices, which can result in Joule heating, this power absorption leads to local heating. Besides heating at the non-resonance conditions, especially at magnetic resonance additional heat will be dissipated into the YIG due to the continuous damping of the YIG magnetization precession.\citep{FrankThermoFMR} The generation of temperature-gradients caused by local heating gives rise to thermoelectric effects such as the Seebeck effect (caused by the difference in Seebeck coefficient of Au and Py), the spin-dependent-Seebeck effect (SdSE) (due to the spin dependency of the Seebeck coefficient in Py, resulting in thermal spin injection at the Au\textbar Py interface), and the spin-Seebeck effect (SSE) (spin pumping caused by thermally excited magnons in the YIG, leading to spin accumulation at the YIG\textbar Au interface).

\begin{figure}[tb]
	\includegraphics[width=8.5cm]{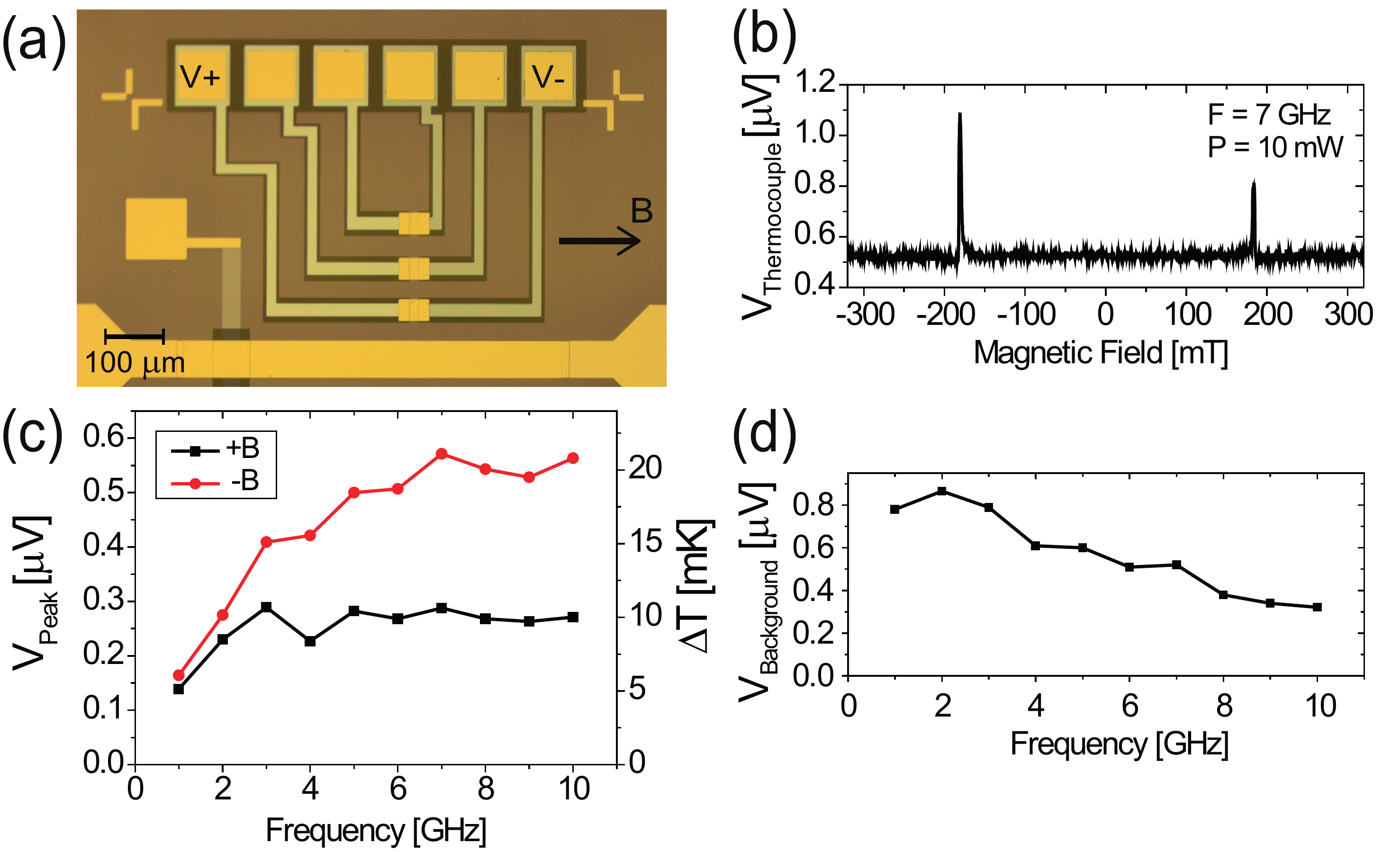}
	\caption{\label{fig:Fig85} 
		(a) Microscope image of the thermo-couples placed near the microstripline. Below all contact-leads a 70-nm-thick Al$_2$O$_3$ layer is present (visible as the dark areas), to avoid spurious signals generated in these leads. Only the Au pad in the center of each thermo-couple is in direct contact to the YIG substrate. Each Au pad is contacted with a 40-nm-thick Pt lead on the left side and a 40-nm-thick NiCu lead on the right side. (b) Detected voltage from the thermocouple most close to the stripline for $F=7$ GHz and $P=10$ mW. (c) Frequency dependence of the detected peaks for positive and negative applied magnetic fields. The right axis shows the corresponding temperature change $\Delta T=\Delta V/(S_\textrm{Pt}-S_\textrm{NiCu})$. (d) Dependence of the thermo-couple background voltage on applied frequency. (c) and (d) are both for $P=10$ mW and for the thermo-couple most close to the stripline.
	}
\end{figure}

In order to probe the RF induced heating, NiCu\textbar Pt thermocouples were placed near the stripline as is shown in Fig. \ref{fig:Fig85}(a). In this way, the temperature of the substrate can locally be measured by making use of the Seebeck effect. From the measured thermo-voltage signals the increase in temperature at the NiCu\textbar Pt junction with respect to the reference temperature of the contact pads can be obtained using $\Delta V = -(S_\textrm{Pt}-S_\textrm{NiCu}) \Delta T$, where $S_\textrm{Pt}=-5$ $\upmu$V/K and $S_\textrm{NiCu}=-32$ $\upmu$V/K are the Seebeck coefficient of Pt and NiCu, respectively.\citep{FrankThermoFMR} Besides a constant background voltage signal, indicating heating when the YIG magnetization is not in resonance, clear peaks are observed at the YIG resonance conditions, as is presented in Fig. \ref{fig:Fig85}(b) for an applied RF frequency of 7 GHz. The magnitude of the peaks at resonance is in the order of 40 nV ($F=1$ GHz, $P=10$ mW, distance from microstrip 195 $\upmu$m) up to 0.6 $\upmu$V ($F=10$ GHz, $P=10$ mW, distance from microstrip 50 $\upmu$m); Higher signals were measured for higher frequencies and for devices closer to the RF line. 

Fig. \ref{fig:Fig85}(c) shows the extracted peak-height of the measurements for the thermocouple most close to the microstrip (50 $\upmu$m), and the corresponding temperature increase is added on the right vertical axis. Additionally, Fig. \ref{fig:Fig85}(d) gives the evolution of the background voltage as a function of applied RF frequency. A maximum temperature increase at resonance conditions of 22 mK is observed in this thermocouple. Interestingly, all measurements show different behavior at the YIG resonance conditions for positive and negative applied magnetic fields; Consistently, the peak at negative applied magnetic fields is larger than the one for positive fields, increasing to a factor 2 in magnitude at $F=10$ GHz. 

These observations are in agreement with the difference in peak-height between positive and negative applied magnetic fields as observed for the Py-devices in area 2, which are placed on the same side of the stripline as the thermocouples. Also here the existence of nonreciprocal magnetostatic surface spin-waves (MSSW) might explain the observed behavior, as they will influence the heating and lead to unidirectional heating of the substrate, as observed by An \textit{et al.},\citep{Unidirectional} who used a measurement configuration very similar to the one we describe in this paper.


The thermocouple measurements show non-negligible heating of the YIG surface, and therefore thermal effects are likely to play a role in the observed voltage generation in the YIG\textbar Au\textbar Py devices. To obtain the quantitative contribution of the Seebeck effect, SdSE, and SSE in the studied device geometry, it is needed to model the YIG\textbar Au\textbar Py device, including its thermal properties. However, from the observed dependence on applied RF frequency of the thermocouple signals, even without knowing the expected quantitative contribution of the thermal effects, it can be concluded that these effects cannot explain the large signals in the YIG\textbar Au\textbar Py devices of area 1. As observed for spin pumping, the thermocouple signals saturate at higher RF frequencies, whereas the series of permalloy devices in area 1 shows a continuously increasing signal. This indicates that additional effects are present, which scale linearly with frequency, such as for example inductive coupling, where the RF current in the microstrip induces a current in the Py, causing Joule heating.

\subsection{Finite element simulation of the SdSE and SSE}

To determine the contribution of the SdSE (at the Au\textbar Py interface) and the SSE (at the YIG\textbar Au interface) we performed a three-dimensional finite element (3D-FEM) simulation of our devices\citep{Comsol} where the charge- ($\vec J$) and heat- ($\vec Q$) current densities are related to the corresponding voltage- ($V$) and temperature- ($T$) gradients using a 3D thermoelectric model. Details of this modeling can be found in Refs. \citep{Comsol, Fasil, bakker_nanoscale_2012, PeltierYIG, FasilSdSE}. The input material parameters, such as the electrical conductivity, thermal conductivity, Seebeck coefficients and Peltier coefficient are adopted from Table I of Refs. \citep{bakker_nanoscale_2012,PeltierYIG}.

\subsubsection{SdSE}

In the SdSE, the heat current flowing across the Au\textbar Py interface causes the injection of spins which are anti-aligned to the magnetization of the Py layer. In our modeling, we use the temperature values measured by the NiCu\textbar Pt thermocouples, shown in Fig. \ref{fig:Fig85}(c), as a Dirichlet boundary condition in the 3D-FEM. 

Specifically, for a given microwave power and frequency, by fixing the temperature of the Au\textbar YIG interface to the measured values and anchoring the leads (Ti\textbar Au contacts) to the reference temperature, we can calculate the resulting temperature gradient $\nabla T_F$ in the Py and hence the SdSE voltage. From this model, for a single Au\textbar Py interface, a total spin-coupled voltage drop of approximately ~$-1$ nV is obtained, which corresponds to  $-96$ nV for the series of Py devices in area 1.

We can also compare this result with one obtained from a simple one-dimensional spin-diffusion model using the following equation \citep{Bram} 
\begin{equation}
\Delta V_s =-2\lambda_s S_{S}\nabla T_{\textrm{Py}} P_{\sigma}R_\textrm{m},\
\end{equation}
where $\lambda_s=5$ nm is the spin-diffusion length, $S_S=-5~ \mu$V/K is the spin-dependent Seebeck coefficient, $P_\sigma=0.3$ is the bulk spin-polarization, $\nabla T_{\text{Py}}=10^5$ K/m is the temperature gradient in the Py, obtained from a simple 1D heat diffusion model across the interfaces, and $R_\text{m}$ is a resistance mismatch term which is a value close to unity for such metallic interfaces considered here. The estimated signal from this 1D-model is a factor two larger than that obtained from the 3D-FEM. Note that, in both modeling schemes, the distance dependence from the strip-line has not been taken into account, which would lead to an even lower signal. Therefore, we conclude that the SdSE does not contribute significantly to the measured signal.




\subsubsection{SSE}
To estimate the maximum contribution of the SSE, caused by spin pumping due to thermal magnons, we need to obtain the temperature difference between the magnons and electrons $\Delta T_\textrm{me}$ at the YIG\textbar Au interface. We again use the 3D-FEM, but this time, extended to include the coupled heat transport by phonons, electrons and magnons with the corresponding heat exchange lengths between each subsystem. The detailed description of this model, which was used earlier to describe the interfacial spin-heat exchange at a Pt\textbar YIG interface, can be found in Ref. \citep{PeltierYIG} along with the used modeling parameters. 

In our model, we set the bottom of the GGG substrate to the surrounding (phonon) temperature $T_0$, the Au\textbar Py interface at the equilibrium temperatures of both electrons and phonons, i.e., $T_\textrm{e}=T_\textrm{ph}=T_0+20$ mK while using a magnon heat conductivity $\kappa_m=0.01$ Wm$^{-1}$K$^{-1}$ and phonon-magnon heat exchange length $\lambda_{\textrm{m-ph}}=1$ nm. From our 3D-FEM model, we find an interface temperature difference of $\Delta T_{\textrm{me}}=25$ $\upmu$K between the magnon and electron subsystems. While $\Delta T_\textrm{me}$ at the YIG\textbar Au interface seem a rather small value, the comparison with earlier reports indicate  equivalence between the ratio of $\Delta T_\textrm{me}$ to the temperature increase of the YIG $\Delta T_\textrm{ph}$.

In the SSE, the spin-current density $J_{s}$ pumped across the YIG\textbar Au interface can be obtained using $J_{s}=L_{S}\Delta T_\textrm{me}$, where $L_{S}=G_r \gamma \hbar k_B/2\pi M_s V_a=7.24\times10^9$ Am$^{-2}$K$^{-1}$ is the interface spin Seebeck coefficient \citep{TheoryMagnon,PeltierYIG} where $G_r$, $M_s$, and $V_a$ are the real part of the spin-mixing conductance per unit area, the YIG saturation magnetization, and the magnetic coherence volume ($\sqrt[3]{V_a} = 1.3$ nm)\citep{SSETheory}, respectively. Because the thickness t of the Au is much smaller than the spin diffusion length $\lambda$ in Au, we can assume a homogeneous spin accumulation $\Delta \mu=J_s t\rho$. The voltage drop at the Au\textbar Py interface is thus $\Delta V_\textrm{SSE}=P_\sigma\Delta \mu/2$, which gives the maximum SSE voltage $V_{\textrm{SSE}}$ detected by the Py. Using $\Delta T_\textrm{me}=25$ $\upmu$K, we obtain $V_{\textrm{SSE}}=54$ pV for a single Py device and a total  of 5 nV for 96 Py serially connected devices. From this discussion we conclude that the combined voltage contributions from the SSE and the SdSE cannot explain the observed enhancement at higher frequencies, suggesting that there are additional effects that need to be considered here.



\subsection{Discussion}
\label{section:discussion}
Besides the measured signals at YIG resonance, a few other present features need attention. First, at low applied frequency and magnetic field, the voltages generated by the Py-devices in series in area 1 show resonance behavior, as is clearly visible in Fig. \ref{fig:Fig82}(e). These resonating signals decrease and finally disappear for higher frequencies. In area 2 also some small resonances are observed [see Fig. \ref{fig:Fig82}(f)], but these resonances are far less prominent. The origin of the resonances might be related to the fact that this system, like the YIG\textbar Pt system, is not only sensitive to the ferromagnetic resonance (FMR) mode, but to any spin-wave mode. This means that additional signals can appear when multiple spin-wave modes exist, which might be more strongly present at lower frequencies, and for some reason more sensitively detected by the narrow strips of Py-devices in series in area 1 as compared to the wider Py-devices and the YIG\textbar Pt system. Furthermore, for the lower frequencies the YIG resonance conditions are very similar to those of the Py layer, which could lead to coupling between those states, resulting in a broader range of possible resonance magnetic fields for a certain applied frequency. 

A second feature that needs attention is the background signal for the YIG\textbar Au\textbar Py devices. The magnitude of the background signal increases with the applied RF power, similar in magnitude as the resonance-peaks, only having opposite sign. From the evolution of the background heating, as depicted in Fig. \ref{fig:Fig85}(d), it is observed that the background heating decreases by increasing the RF frequency. Therefore it is not possible to directly attribute the measured background signals of the YIG\textbar Au\textbar Py devices to heating of the substrate, and the origin of these signals is still unclear.

Third are the peaks observed at Py resonance conditions. As a control experiment the same sequence of YIG\textbar Au\textbar Py devices was fabricated on a Si\textbar SiO$_2$ substrate, including the waveguide and stripline. In these samples no resonance peaks were present, neither at the YIG resonance conditions nor at the Py resonance conditions. This experiment proves the need of the YIG substrate in order to detect Py resonance.

A possible explanation of the origin of the detected peaks is as follows: By having the Py magnetization in resonance, a pure spin-current is pumped into the adjacent layers. The polarization of this spin-current consists of both an ac- and a dc-component. The spin-current pumped into the upper contact will relax and does not give rise to any signal. The spin-current pumped into the thin Au layer below the Py strip will not relax before arriving at the YIG\textbar Au interface. At this interface the component of the spin angular momentum perpendicular to the YIG-magnetization (here the ac-component of the pumped spins) will be absorbed and the parallel component (the dc-component of the pumped spins) will be reflected (as is the case for the spin-Hall magnetoresistance). This interaction with the YIG\textbar Au interface results in only the dc-component of the initially pumped spin-current being reflected. The reflected spins will diffuse back to the Py strip, where they accumulate, as their polarization direction is changed as compared to the spins being pumped. This spin-accumulation results in a build-up potential, which is measured. In the case YIG is replaced by SiO$_2$ this mechanism does not work, as the absorption and reflection of spins at the SiO$_2$\textbar Au interface is not spin-dependent.



\begin{figure}[]
	\includegraphics[width=8.5cm]{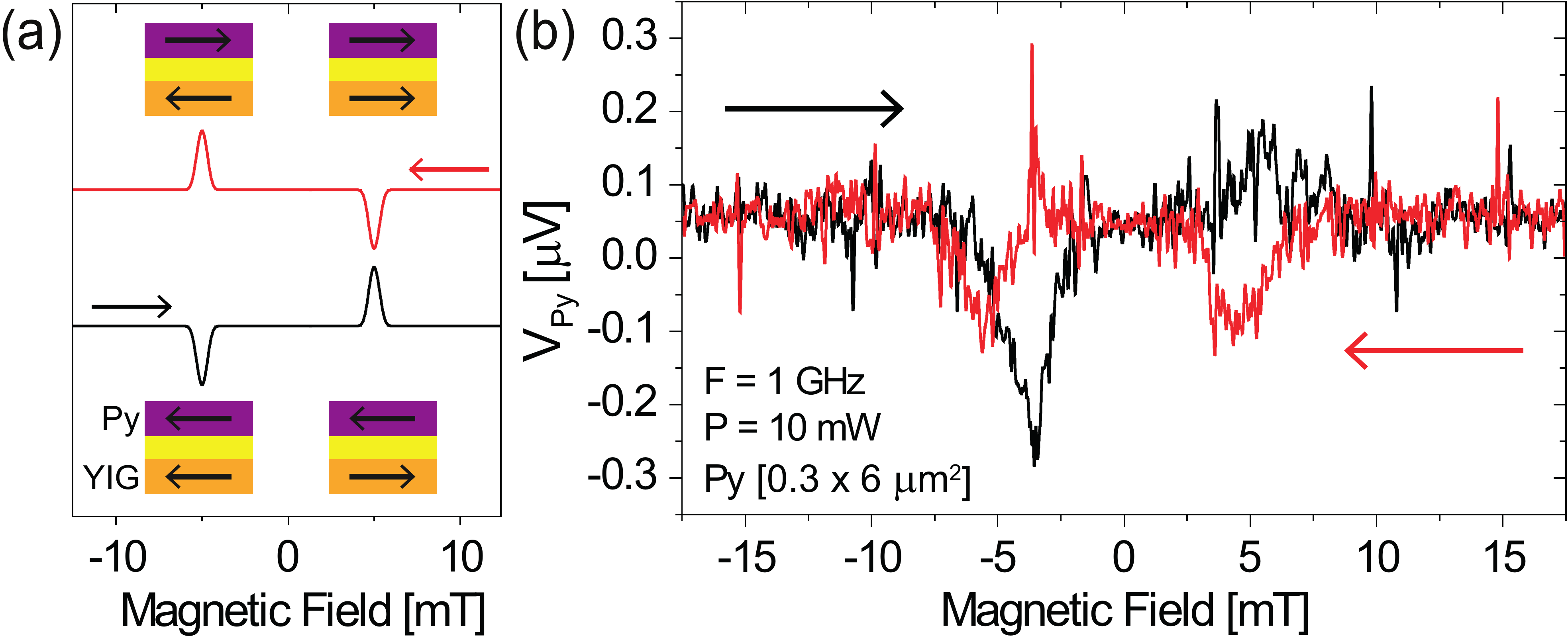}
	\caption{\label{fig:Fig86} 
		(a) Theoretically expected resonance peaks, when individual switching of the YIG and Py layers is obtained. The insets show the magnetization orientation of the YIG and Py layers for each resonance peak. (b) Measurement result for a batch of samples having smaller Py-strips ($0.3 \times 6$ $\upmu$m$^2$) as compared to the afore described devices, in order to observe the anti-parallel state of the YIG and Py magnetization directions. The trace and retrace of the measurement are marked by the black and red data, respectively. 
	}
\end{figure}

Finally, to prove the obtained signals are caused by spin pumping, and detected by the diffusion of the generated spin accumulation to the Py layers, as in typical non-local spin-valve devices, the observed peaks should change sign when the magnetization directions of the YIG and Py are placed anti-parallel, as is shown in Fig. \ref{fig:Fig86}(a). In the experiments, this situation turned out to be hard to accomplish, as the switching fields of both YIG and Py are relatively low: for YIG smaller than 1 mT, and for the 20-nm-thick Py strips on YIG a maximum switching field of 10 mT was obtained for Py dimensions of $0.3 \times 6$ $\upmu$m$^2$. So when sweeping the magnetic field, only fields between 1 and 10 mT will result in anti-parallel alignment of the magnetic layers. Besides this field window being rather narrow, the resonance peaks in this regime are no clear single peaks [see Fig. \ref{fig:Fig82}(e)]. Nevetheless, one batch of samples was fabricated where the Py dimensions were set to $0.3 \times 6$ $\upmu$m$^2$, and one resulting measurement is shown in Fig. \ref{fig:Fig86}(b). While sweeping the magnetic field from negative to positive values (black line), the resonance peak clearly changes sign. However, for the reverse field sweep (red line) the expected switching of the peak is not observed: There is a narrow positive peak, but less than halfway the expected peakwidth, it reverses sign. From this measurement it is not possible to unambiguously state the presence of the sign reversal for the anti-parallel state. To do so, a device is needed having a switching field of the second magnetic layer in the order of 100 mT or higher (possibly accomplished by replacing Py for cobalt, which has a larger coercive field), such that the anti-parallel magnetization state can also be obtained for slightly higher magnetic fields.

\section{Summary}
In summary, we have observed the generation of voltage signals in YIG\textbar Au\textbar Py devices placed in series, caused by YIG magnetization resonance. Furthermore, the resonance of the magnetic Py layers, caused by direct excitation of the magnetization, or indirect dynamic coupling, is detected. By modeling our device structure, we find that the signals of the wider Py structures (area 2) can very well be reproduced by the calculated spin pumping signals. For the narrow structures (area 1) additional signals are detected. The origin of these additionally observed signals and some other features, such as the dependence of the resonance peaks on applied RF frequency, the resonating peaks at low applied frequencies, and the increasing background voltage as a function of RF frequency, remain to be explained. 

Spin-dependent thermal effect are also quantified; The heating caused by the applied RF current was studied by placing thermocouples in close proximity to the stripline. Due to the temperature increase at the surface of the YIG substrate, especially at YIG resonance conditions, contributions of thermal effects to the generated voltage in the YIG\textbar Au\textbar Py devices cannot be excluded. Nevertheless, from finite element simulations we find that the contribution of the SdSE and SSE are rather small and cannot explain the observed features. Additionally, the thermocouples showed that the heating of the substrate is dependent on the applied field direction, indicating the possible presence of nonreciprocal magnetostatic surface spin-wave modes.

Concluding, we have shown the possibility to electrically detect magnetization resonance from an electrical insulating material by a spin-valve-like structure, without making use of the ISHE. For the presented work, 96 and 62 Au\textbar Py devices were placed in series to increase the magnitude of the generated signal. By comparing the obtained data with signals from a reference Pt strip and using a finite element model of the devices, we find that part of the detected signals can be ascribed to spin-current generation by spin pumping (for the 62 devices in area 2 an agreement between measurements and calculated signals within 20\% is found), however, especially for the 96 devices in area 1, additional signals are present, of which the origin remains to be explained. Once a better understanding of the origin of the full signals is obtained, the device geometry and injection efficiency can be improved, such that the number of needed devices can be decreased, which opens up possibilities for new types of spintronic devices, where magnetic insulators can be integrated.


\section*{Acknowledgements}
We would like to acknowledge J. Flipse for sharing his ideas and M. de Roosz, H. Adema and J. G. Holstein for technical assistance. This work is supported by NanoNextNL, a micro and nanotechnology consortium of the Government of the Netherlands and 130 partners, by NanoLab NL and by the Zernike Institute for Advanced Materials (Dieptestrategie program).

\bibliography{YIGPt}

\end{document}